\documentstyle[aps,preprint,eqsecnum,epsfig,rotate]{revtex}
\tightenlines
\begin{document}
\title{Interquark potential, susceptibilities and particle density  of \\
two color QCD at finite chemical potential and temperature}
\author
{Maria-Paola Lombardo}
\address{Istituto Nazionale di Fisica Nucleare \\
Laboratori Nazionali del Gran Sasso \\ 
S.S. 17 bis Km. 18,910 I-67010 Assergi (AQ), Italy \\ 
lombardo@lngs.infn.it }
\maketitle

\begin{abstract}

We explore the phase diagram of  SU(2) Lattice Gauge Theory with 
dynamical fermions in the temperature, mass, chemical potential space. 
We observe qualitative changes of the dependence of the particle
density  on $\mu$ and $T$, which is compatible with that expected
of a gas of free massless quarks $n \propto \mu^3$ only for $T \simeq T_c$.
At the onset for thermodynamics the interquark potential flattens at 
large separations, indicating enhanced fermion screening and the 
transition to a deconfined
phase. Temporal and spatial Polyakov loops  behave in different
ways, the latter being nearly insensitive to the chemical potential. 
The rotation of the chiral condensate to a baryonic condensate,
as inferred from the susceptibilities,  might occur together with a 
reduction of its magnitude in the chiral limit, 
possibly leading to a critical  temperature 
for  diquark  condensation smaller than the deconfinement  temperature. 
We further asses the r\^ole of the chemical potential into the gauge
dynamics by carrying out a partial quenched calculation. 
We speculate on the relevance, or lack thereof, of our findings to real QCD.

\end{abstract}

\section {Introduction}

A so far unique possibility to
study a gauge model at finite density is afforded by
two color QCD, which can be studied via numerical simulations even
at non--zero chemical potential. 
In our first paper \cite{HKLM} 
(hereafter HKLM)  we have studied the model  at zero temperature,
presenting the first {\em ab initio} spectrum calculation
at finite baryon  density with full fermion feedback. 
Here we focus on the effect of temperature: we 
discuss the interplay of bulk thermodynamics observables, deconfinement
and chiral symmetry by varying the chemical potential
at different temperatures, and we contrast the nature of the 
phenomena along the temperature and density axes.
Ultimately, we would like to understand the phase diagram of
two color QCD, the nature of the
critical lines separating the phases with different realisation of
chiral symmetries and their interplay with the
deconfinement and topological transitions. 
Finally, and ideally, we would also like to use some of the 
results on two color QCD to learn something about three color QCD.
Clearly the present study, which addresses only a subset of these issues
on a small $6^4$ lattice is just exploratory :
we stress that even the ``simple'' $\mu=0$ limit, whose study was 
initiated many years ago in \cite{john}, has a number of  interesting open 
problems\cite{su2varie}. 

A rich pheonomenology at finite baryon density emerges
from the analysis of model Hamiltonians with the same 
global symmetries as QCD\cite{mod}.
For two colors the predictions of these more sophisticated studies
reduce to those expected of qualitative reasoning and early
calculations \cite{diqcon}. 
Simple models and qualitative reasoning, clearly, do not 
take fully into account details of the dynamics such as
color forces and their temperature dependence:
perhaps some of the lattice results  can offer some quantitative control, 
and further insight into model studies,
and, very optimistically, even the possibility of a
first principle derivation
of some of their relevant parameters (e.g. form factors).
This provides a further motivation for this lattice study of two color
QCD.

An overview of the parameter space explored by the numerical simulations
is given in Section II, while the
results for the quark-(anti)quark potential are presented in 
Section III. Section
IV discusses the particle number, and the ``nuclear matter phase''. 
Section V presents the data for
the susceptibilities, contrasting the high density and high temperature
behaviour.
In Section VI we further assess the r\^ole of the chemical potential 
into the gauge dynamics by carrying out a partial quenched 
calculation, and present some speculative comments on the relationships of
two and three color QCD. We close with
a summing up and outlook. Some of the results of this paper have
been briefly mentioned in \cite{ect}.

\section{Overview of the parameter space}

Using the same algorithm as in HKLM  we studied the
thermodynamics of the system on a $6^4$ lattice at:
\begin{itemize}
\item $\beta = 1.3$, $\mu = (0, .2, .6, .8)$, mass = (.07, .05);
\item $\beta = 1.5$, $\mu = (0, .2, .6, .8, 1.0)$, mass = (.1, .07, .05); 
\item $\mu=0$, $\beta = (1.3, 1.5, 1.7, 1.9, 2.1, 2.3)$, mass = .1.
\end{itemize}
The chemical potential explores the range of interest, up to the lattice 
saturation, while the temperature ranges from $T \simeq 0$ to $T> T_c$.
By contrasting the results obtained at $\beta = 1.5$ with those
obtained at $\beta = 1.3$ we shall study how the temperature effects the
chemical potential dependence. By contrasting the results
as a function of chemical potential with those obtained by varying
the temperature we shall compare the nature of the high T and the high
$\mu$ transitions. Wherever available we rely on
the results for condensates and susceptibilities  obtained
in HKLM.

Clearly to study the chemical potential dependence at different temperature
we need  to know in which phase we are in at $\mu=0$.
Our first task is then (approximately) locate the critical line 
in the $\beta$--mass plane. 
To this end we measured the mass dependence of the chiral condensate. 

\begin{figure}
\begin{center}
{
\epsfig{file=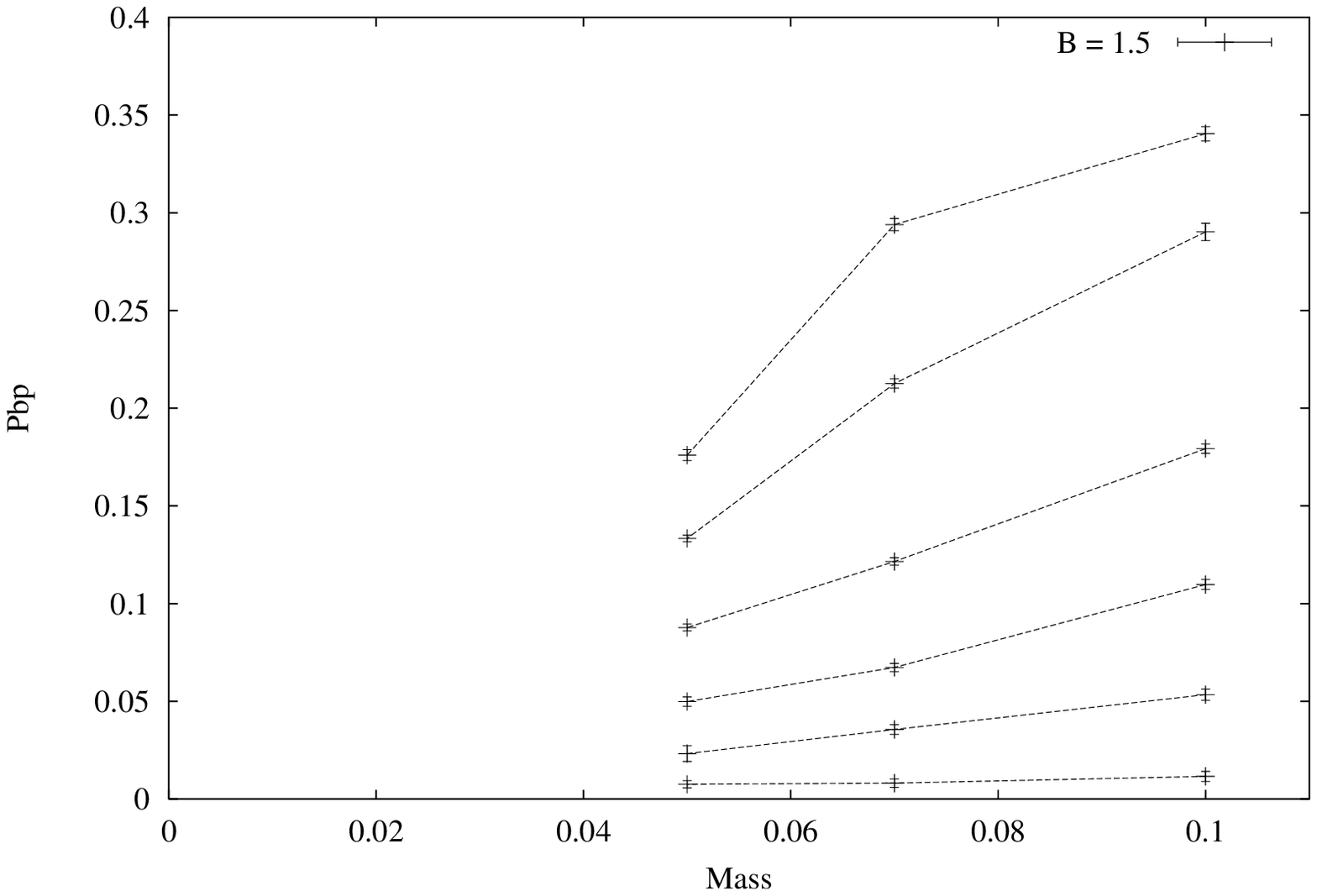, width=12 truecm}}
\end{center}
\caption[xxx]{$<\bar\psi\psi>$ as a function of the bare mass,
 $\beta = 1.5$ 
 $\mu = 0, .2, .4, .6, .8, 1.0 $ 
from top to bottom} \label{fig:pbp_vs_mass}
\end{figure}

\begin{figure}
\begin{center}
{\epsfig{file=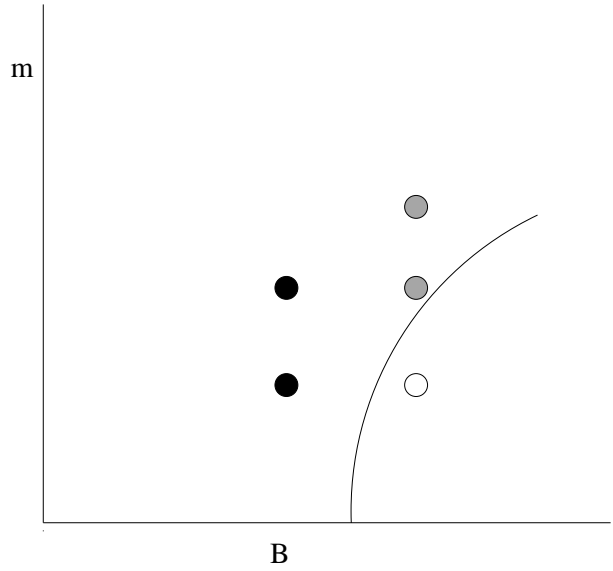, width=8 truecm}}
\end{center}
\caption[xxx]{The critical line in the $\beta$--mass plane and the
simulation points at $\beta = (1.3, 1.5)$, mass=(.05, .07, .1)}
\label{fig:m_b}
\end{figure}

In Fig. \ref{fig:pbp_vs_mass}
we show the chiral condensate as
a function of the bare mass for different values of the 
chemical potential at $\beta = 1.5$.
We note that, while a linear extrapolation from 
masses .1 and .07 would give a non--zero condensate, an
extrapolation which uses masses .07 and .05 
would suggest that chiral symmetry is 
already restored. In addition to that,  
for  mass = .05 we observed a clear two state signal in the HMD history
of the chiral condensate, which, to a lesser extent, was also visible at 
mass = .07.  Fig. \ref{fig:pbp_vs_mass} 
should be considered together with the analogous
one ( Fig. 3) of HKLM from which we inferred that the
points $\beta = 1.3$,  mass = .05 and .07 are in the phase 
where $<\bar \psi \psi> \ne 0$.
Assuming a first order chiral transition in $\beta$ at $m=0$, 
we would conclude that the chiral transition line in the $\beta-m$ plane
runs close to  the points $(\beta, m)  = (1.5, .05), (1.5, .07)$.
We sketch such a line in Fig. \ref{fig:m_b}, with the caveat that
if the $m=0$ transition were instead second order, the 
line drawn in Fig. \ref{fig:m_b} would indicate a crossover. 
Given the exploratory nature of this study this distinction
is immaterial. In the same Figure we also show the locations of the simulation 
points $\beta = (1.3, 1.5)$, and $m = (.05, .07, .1)$:
by switching on the chemical potential we should be able to observe
the effect of a finite density of quarks on a variety of dynamical
situations, ranging from ``very cold'' (filled points), 
to ``critical'' (shaded points) to ``hot'' (open point). In addition
to this, the line $m = 0.1$ has been explored at $\mu=0$ for several 
$\beta's$.

\begin{figure}
\begin{center}
{\epsfig{file=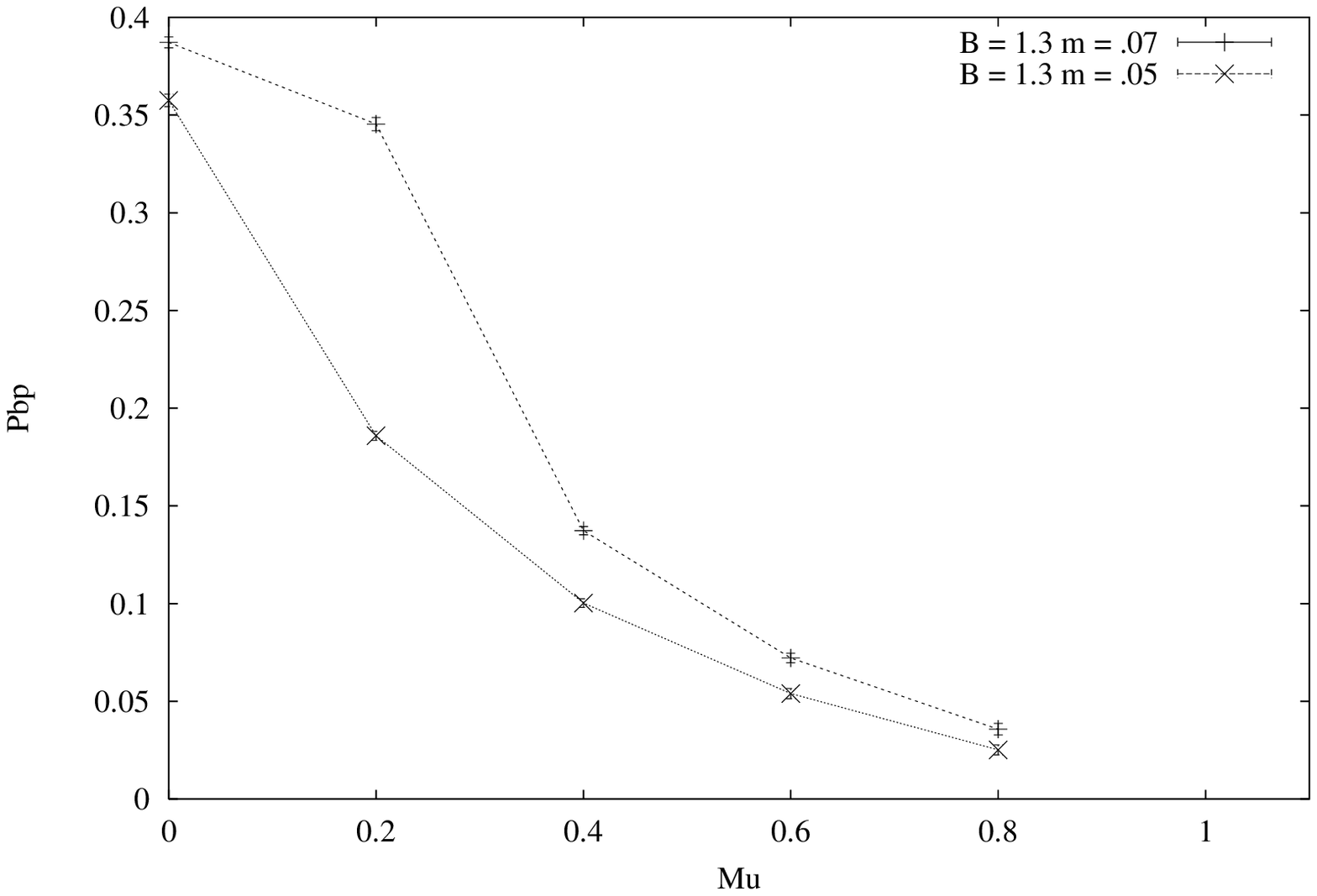, width= 12 truecm} \\
\epsfig{file=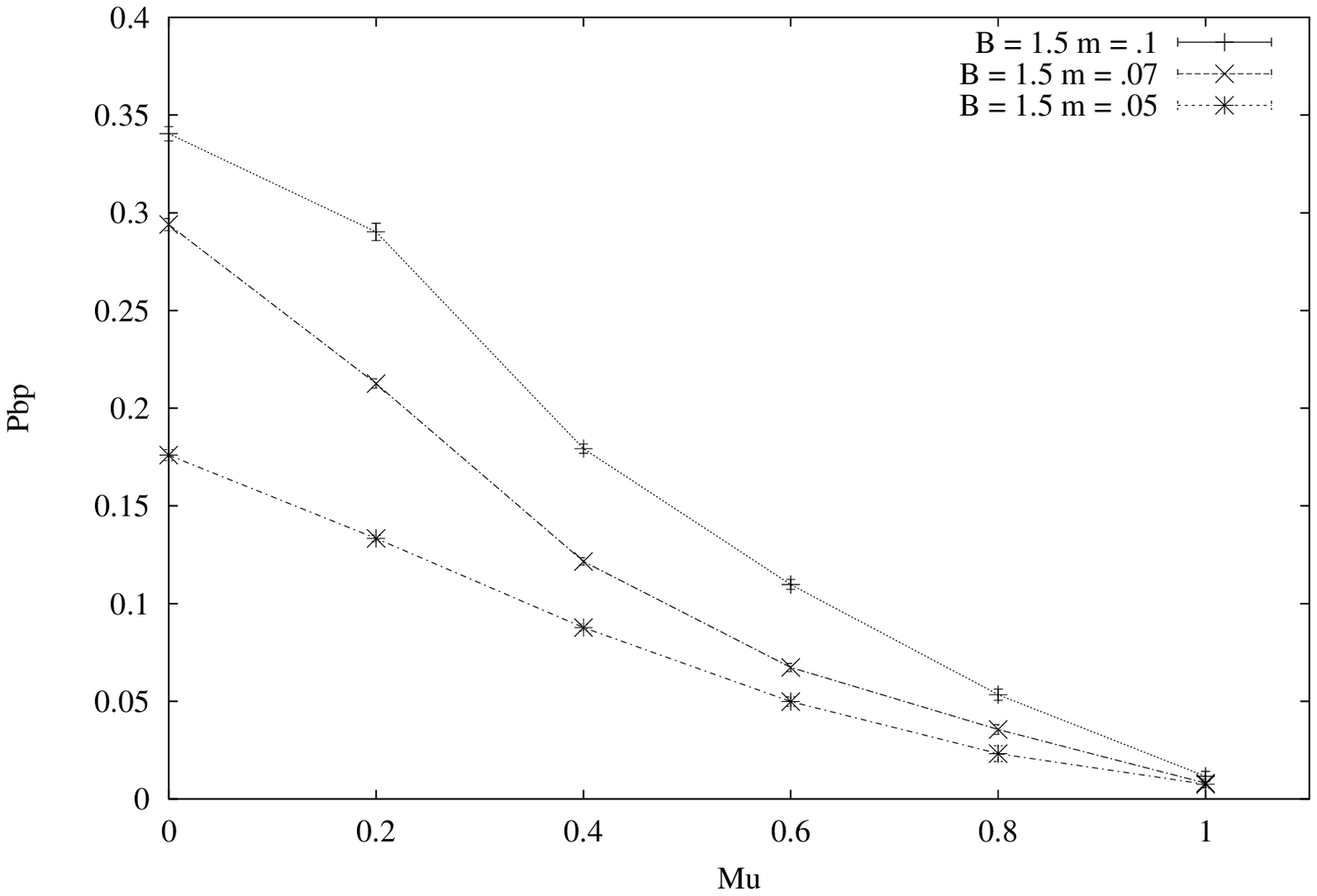, width= 12 truecm}}
\end{center}
\caption[xxx]{$<\bar\psi\psi>$ as a function of the 
chemical potential for $\beta = 1.3$ (upper, reproduced from HKLM) 
and $\beta = 1.5$,  and masses as shown in the diagrams}
\label{fig:pbp_vs_mu}
\end{figure}

In Fig. \ref{fig:pbp_vs_mass} it is also shown the dependence of the chiral
condensate on the chemical potential. As expected, when $\mu$ is increased 
the chiral condensate extrapolates to zero in the chiral limit also for 
larger bare masses. We remind that, of course, this does not correspond
to the restoration of the chiral symmetry as the chiral condensate will be
rotated to a baryonic diquark one 
( see the discussions in HKLM, and below, for more).
Figs. \ref{fig:pbp_vs_mu}
show the same $<\bar \psi \psi>$ data 
plotted as a function of the chemical
potential. Again,  the findings at this largest temperature should
be contrasted with those obtained in HKLM
which we reproduce for the reader's convenience in the upper part
of the figure. 
At $\beta = 1.3$, mass = .07 we observe
a rather sharp drop in the chiral condensate at $\mu \simeq .3$,
close to the half the pion mass, $m_\pi/2 \simeq .3$. 
We also note that at $\beta = 1.5$,  $m = .1$ 
$m_\pi= .8$,  and $\mu \simeq m_\pi/2 \simeq .4 $ lies somewhere in 
the crossover region as it should.

The overall observation resulting from Fig.  \ref{fig:pbp_vs_mass} is
that the density transition shows the expected softening 
at smaller masses and  higher temperatures.

\section{Quark-antiquark and quark-quark interactions }

Although the results for $<\bar \psi \psi>$ look physical, and
consistent with expectations, we
would have obtained similar ones by use of the quenched
approximation: we are not really observing qualitative, 
dynamical $\mu$ effects. It is interesting to investigate
the dynamics of the gauge fields, and to this end we measured 
the correlations 
$ <P(O) P^\dagger (z)>$ of
the zero momentum Polyakov loops, averaged over spatial directions.
Remember that
this quantity is related to the string tension $\sigma$ via
$\lim_{z \to \infty}  <P(O) P^\dagger(z)> \propto e^{-\sigma z}$.

Note that as $P$ is a real quantity  the Polyakov loop correlator
defining the strength of the quark-quark
interaction $ <P(O) P (z)> / < P(0)^2>$ 
would be the same as the quark--antiquark one
$ <P(O) P^\dagger (z)/ < |P(0)|^2>$.
This would lead to the conclusion that the diquarks and mesons
with the same spin and opposite parity are always degenerate, since their
binding forces are the same. In principle this conclusion is limited to
the heavy spectrum,
however the numerical results presented in HKML support this point
of view: mesons and diquarks seem to remain degenerate in SU(2) at
any nonzero density, once the Fermi level shift is taken into account,
producing four degenerate particles at large $\mu$.

\begin{figure}
\begin{center}
{\epsfig{file=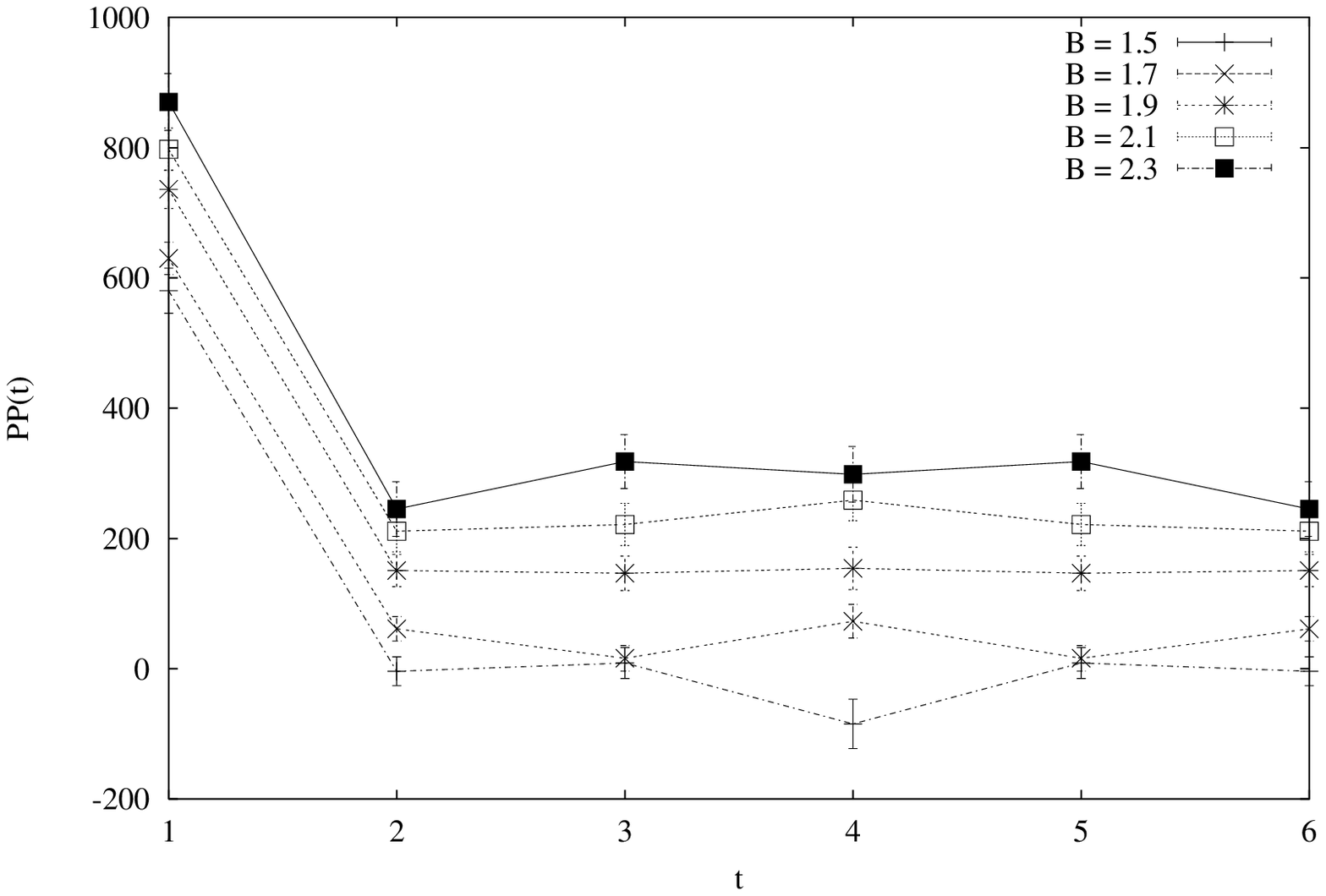, width = 12 truecm} \\
\epsfig{file=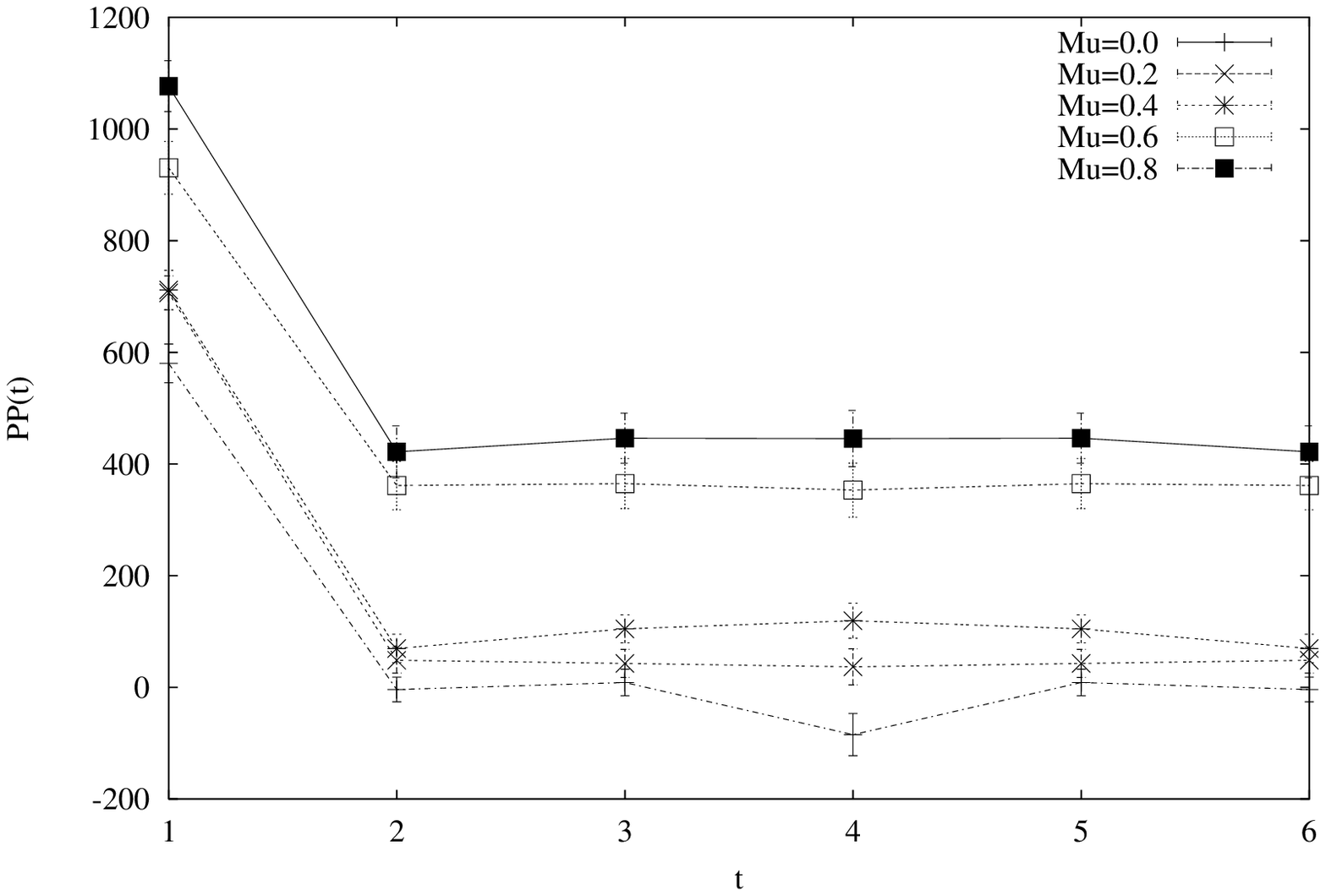, width = 12 truecm}}
\end{center}
\caption[xxx]{Correlations of the zero momentum Polyakov loop as
a function of the space separation. The upper diagram is for
$\mu = 0$, and $\beta$ as indicated. The lower part is for $\beta =
1.5$ and $\mu$ as indicated. In both cases we observe long
range ordering possibly associated with deconfinement}
\label{fig:polcor}
\end{figure}

We show the results for the Polyakov loop
correlators in Fig. \ref{fig:polcor}, where we compare the
behaviour at various temperature (upper) with that at various
chemical potentials (lower). We note that the trends with temperature and 
chemical potential  are quite similar: in both cases
we have signs of long range ordering, i.e. deconfinement. The gap between
the plateaus having $\mu=.4$ and $\mu=.6$ in Fig. 3b suggests increased fermion
screening and the passage to a deconfined phase. We have
then a direct evidence of the effect of the chemical potential
on the gauge fields. 

\begin{figure}
\begin{center}
{\epsfig{file=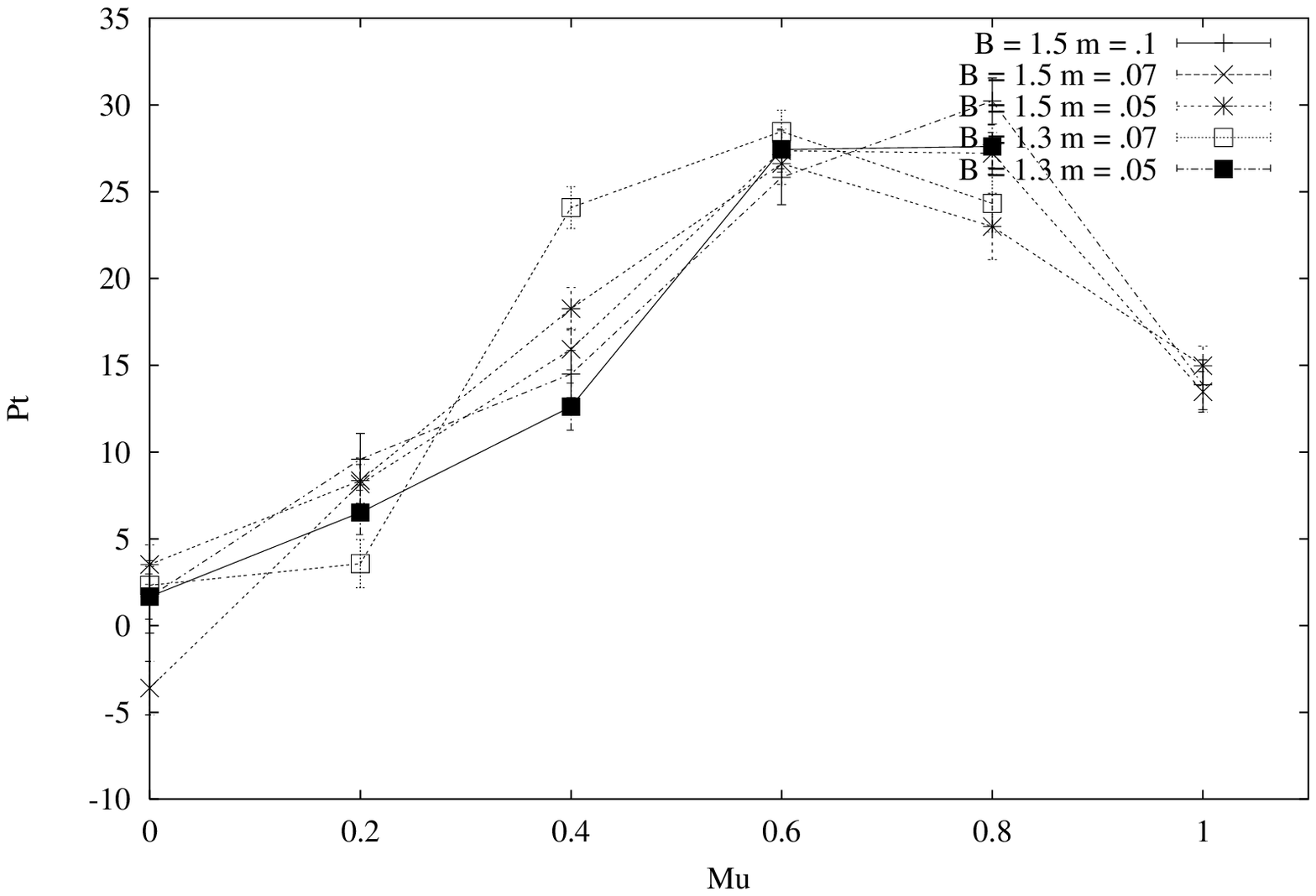, width = 12 truecm} \\
\epsfig{file=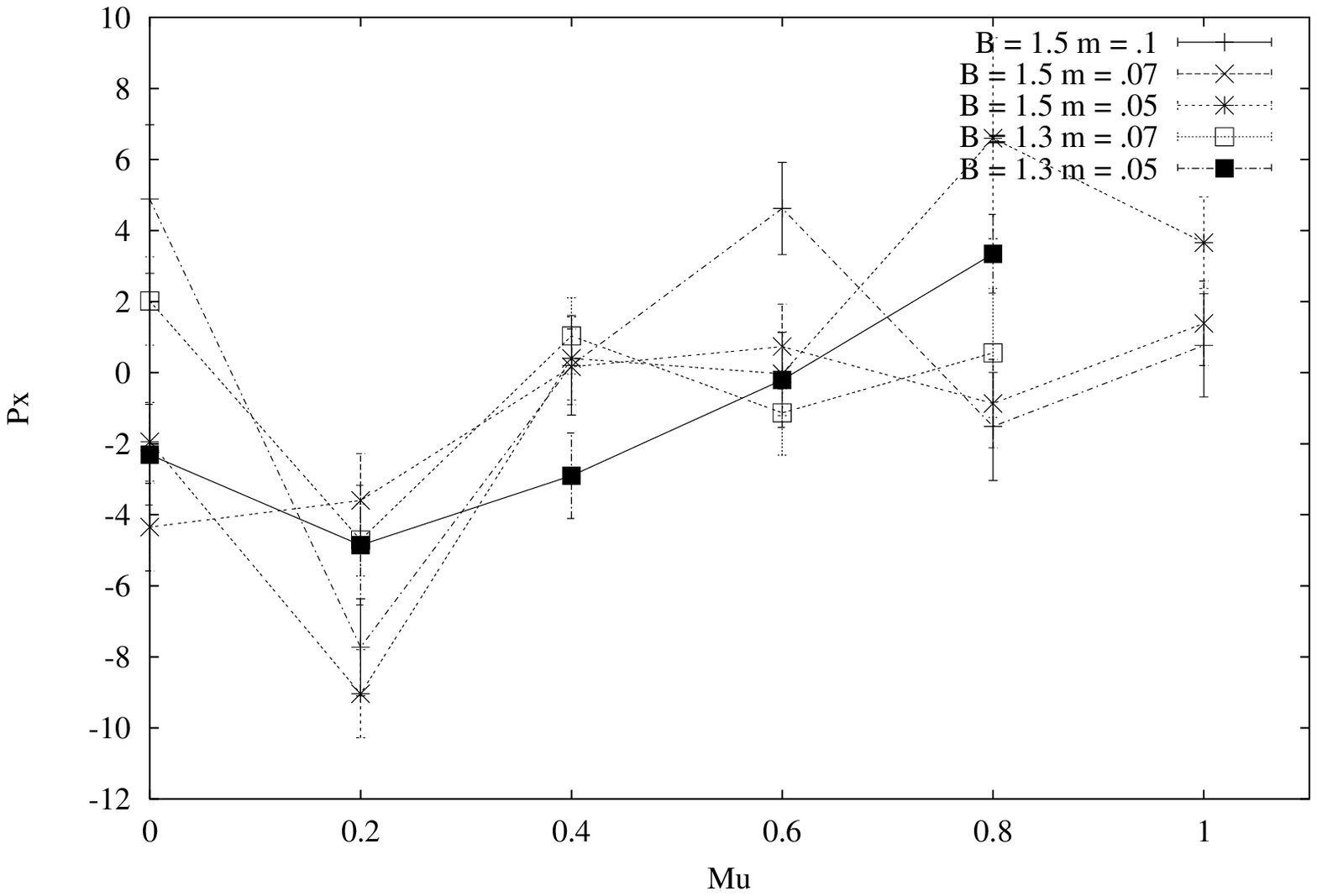, width = 12 truecm}}
\end{center}
\caption[xxx]{Polyakov loops as
a function of the chemical potential. The upper diagram is for
the temporal Polyakov loops, the lower diagram for one spacial
direction (the others behave similarly)}
\label{fig:poltot}
\end{figure}

Finally, we show (Fig. \ref{fig:poltot}) the dependence of the Polyakov 
loop itself on the chemical potential, contrasted, in the lower
part of the figure, with that of the spatial Polyakov loop. 
This suggests that the ordering effects of the chemical potential
only affects the temporal direction. We should also note that
some distinction between the behaviour of
temporal and spacial Polyakov loop is also seen in the temperature
dependence, and can be tentatively ascribed to an effect of the
different boundaries for the fermion fields (periodic vs. antiperiodic)
in space and time direction.

\section{The nuclear matter of SU(2), and a ``new vacuum''?}

According to the standard wisdom for a chemical potential 
comparable with the baryon mass, baryons start to be produced 
thus originating
a phase of cold, dense matter. For SU(2) baryons (diquarks) 
are bosons (as opposed to the fermionic  baryons of real
QCD). This has major implications on the physics of the dense
phase. First, and obviously, the thermodynamics of (interacting)
Bose and Fermi gases is different.  In particular, 
it could be that
the energetics favours the condensations of bosons. This would
be revealed by a non--zero diquark condensate
\cite{diqcon}, which 
still breaks chiral symmetry, and should be reflected by different
functional dependence of the particle density
on $\mu$ which can thus characterise the various phases.

\begin{figure}
\begin{center}
{\epsfig{file=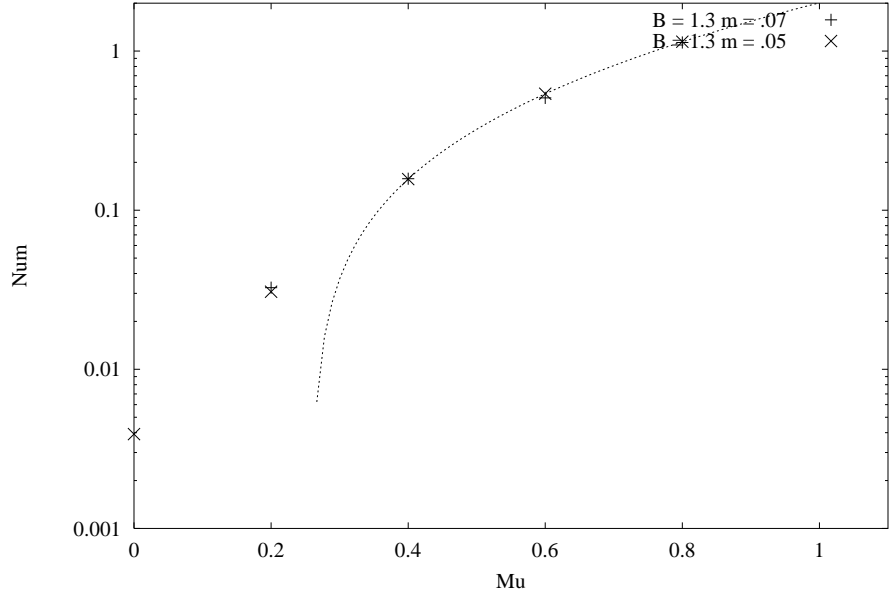, width = 12 truecm} \\
\epsfig{file=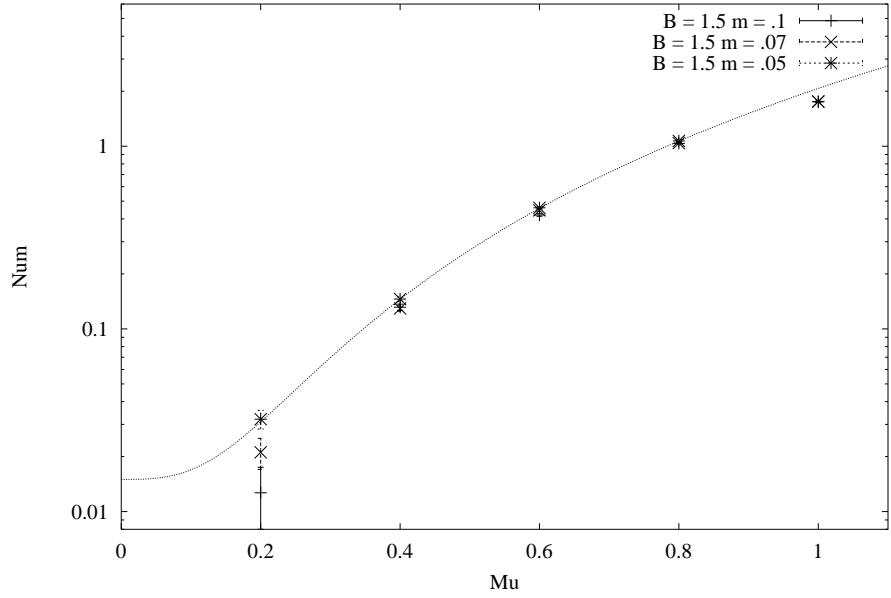, width = 12 truecm}}
\end{center}
    \caption[xxx]{Number density as a function of
the chemical potential, for different masses. The cubic fits
to $a\mu^3 + b\mu^2 +c$ are superimposed. b, c $\simeq 0$ for
$\beta = 1.5$ : the behaviour is then consistent with a pure quark gas.} 
\label{fig:lognum_vs_mu}
\end{figure}

In this exploratory study we just (try to) fit the data
to a cubic spline, appropriate from free massless fermion at zero temperature,
and try to infer from that the effect of temperature, 
and the possibility of a condensed diquark phase. 
In Figs. \ref {fig:lognum_vs_mu}  we show the number
density as a function of $\mu$ for the two $\beta's$ values,
with superimposed the fits described below.

Let us consider $\beta = 1.5$ (our warmer lattice, close
to $T_c$) first: we note the following features.
At small chemical potential there is a clear dependence on the bare
mass, which is greatly reduced in the thermodynamical region. This
could be ascribed to a loss of the dynamical mass. In the same region,
the behaviour is close to $\mu^3$ (the line connecting the point is
the polynomial $2.07 \mu^3 - .01 \mu^2 + .01$). We just
note here that a simple, pure cubic term accommodates the data well
suggesting a free quark gas --
somewhat surprisingly, the contribution of the temperature term seems 
to be small  \cite {thanks}.
At $\beta = 1.3$ the situation is different. For $\mu > \mu_c$,
we would expect diquark condensation. At the same
time, we have observed long distance screening and deconfinement, so
the behaviour might get closer to that of a cold phase of free quarks, 
which, however, is not supported by the data :
as the diquark states appear to be bound (HKLM),
their condensates can still influence the thermodynamics.
The polynomial fits   $ n(\mu) = .84 \mu^3 +1.27 \mu^2 - .1$ for mass = .05,
 $ n(\mu) = 1.2 \mu^3 + 0.5 \mu^2 - .03$ for mass = .07 are not
amenable to any simple intepretation. We might well expect a
rather complicated ``mixed'' nuclear matter phase, with an
admixtures of diquark gas (which, if alone, would constitute a
``pure'' SU(2) nuclear matter), condensed diquarks, free quarks, perhaps
characterised by both types of condensates (such mixed phases are
also predicted by more detailed instanton studies \cite{new}). 
A direct measure of diquark condensate should completely clarify this point
\cite{sisu}. A free massless quark phase, 
with complete  restoration of chiral symmetry 
(i.e. $<\bar \psi \psi> = <\psi \psi> = 0$) could be reached  at even 
larger $\mu$ --  as this region is dominated by lattice saturation artifacts,  
improved/perfect actions \cite{wuwe} might be necessary to explore it.

\section{Particle content and condensation}

Clearly to better understand the thermodynamics results shown above
it is useful to assess the particle content of the model, together
with their possible condensation. Remember that at $\mu=0$ the model
has a lattice analogous of the 
Pauli--G\"ursey symmetry mixing quarks and antiquarks which belong
to equivalent representation. Main consequences are 1. the only
discernable condensate is $<qq>^2 + <q\bar q>^2$, 2. the preferred direction
of $\chi SB$ is picked by the explicit mass term 3. the diquark condensate
is a natural object which does not break any more symmetry than the
ordinary condensate 4. there are massless baryons (diquarks). When
$\mu \ne 0$ the symmetry group is smaller, coincident
with that of staggered fermions, and the number of Goldstone modes  is 
reduced accordingly. 

We refer to HKLM for more detailed discussions, and just
reproduce here the 
basic idea and definitions used in spectroscopy calculations. 

As usual  we form meson and diquark 
operators by taking  correlations of the quark propagator in the appropriate
sector of quantum numbers. We shall limit ourselves to the local sector
of the spectrum and focus on the zero momentum {\it connected} propagators
of the scalar and
psudoscalar mesons and diquarks. The scalar meson propagator
will thus be an isovector, which we will call $\delta$, following QCD
notation.

By applying 
a generic O(2f) transformation to the  mass term
$\bar \chi \chi$ we identify the basic set of operators which
shall be used to build the spectrum:

\begin{eqnarray}
\mbox{\rm scalar}\qquad\chi_1 \bar \chi_1 + \chi_2 \bar \chi_2 &\hskip 1 truecm 
&  
\mbox{\rm pseudoscalar}\qquad
\varepsilon(\chi_1 \bar \chi_1 + \chi_2 \bar \chi_2) 
\\
\mbox{\rm scalar diquark }\qquad\chi_1 \chi_2 - \chi_2 \chi_1 & \hskip 1 truecm& 
\mbox{\rm pseudoscalar diquark}\qquad
\varepsilon(\chi_1 \chi_2 -  \chi_2 \chi_1) \\
\mbox{\rm scalar antidiquark}\qquad\bar \chi_1 \bar \chi_2 - \bar \chi_2 \bar 
\chi_1  & \hskip 1 truecm&
\mbox{\rm pseudoscalar antidiquark}\qquad
\varepsilon (\bar \chi_1 \bar \chi_2 -\bar 
\chi_2 \bar \chi_1)
\end{eqnarray}
where the lower index labels colour.  The first line displays
the usual pseudoscalar and scalar operators. The second (third) line
corresponds to diquark (antiquark) operators, scalar and pseudoscalar. 
This simple minded quantum number assignment can be done by considering
that  quark -- quark and quark-antiquark pairs have
opposite relative parity, and is confirmed by a more rigorous analysis
 presented in HKLM.

Consider now quark propagation from a source at 0 to the point $x$.
The propagator $G_{ij}$ ($i,j$ color index) is an SU(2) matrix:

\begin{equation}
G_{ij} = \left(\matrix {a & b \cr
                   -b^\star & a^\star} \right)
\end{equation}
  
The meson ($q\bar q$) and diquark ($qq$) and antidiquark ($\bar{q}\bar{q}$) 
propagators at 
$\mu=0$ are constructed from $G_{ij}$
as follows:-

\begin{eqnarray}
\mbox{\rm pion}\qquad \mbox{tr}GG^{\dagger}&=&  (a^2 + b^2) \label{eq:Gpion}\\
\mbox{\rm scalar meson}\qquad \varepsilon\mbox{tr} GG^{\dagger}
& =& \varepsilon (a^2 + b^2)\\
\mbox{\rm scalar $qq$}\qquad \mbox{det} G& =&  (a^2 + b^2) 
\label{eq:scalarqq}\\
\mbox{\rm scalar $\bar q\bar q$}\qquad  \mbox{det} G^{\dagger}& =& 
 (a^2 + b^2) \\
\mbox{\rm pseudoscalar $qq$}\qquad \varepsilon \mbox{det} G& =&
 \varepsilon (a^2 + b^2) \\
\mbox{\rm pseudoscalar $\bar q\bar q$}\qquad \varepsilon \mbox{det} 
G^\dagger
& =& \varepsilon (a^2 + b^2)
\label{eq:pseudoqq}
\end{eqnarray}

The notable feature of the propagators at $\mu=0$ is the exact degeneracies of
the pion, scalar $qq$ and scalar $\bar{q}\bar{q}$ and of the scalar meson,
pseudoscalar $qq$ and pseudoscalar $\bar{q}\bar{q}$. 
We then identify two orthogonal directions in the chiral space:\\
a) $\pi$ -scalar diquark - scalar antidiquark \\
b) $\delta$ - pseudoscalar diquark - pseudoscalar antidiquark

All this at $T=\mu=0$.

\begin{figure}
{\epsfig{file=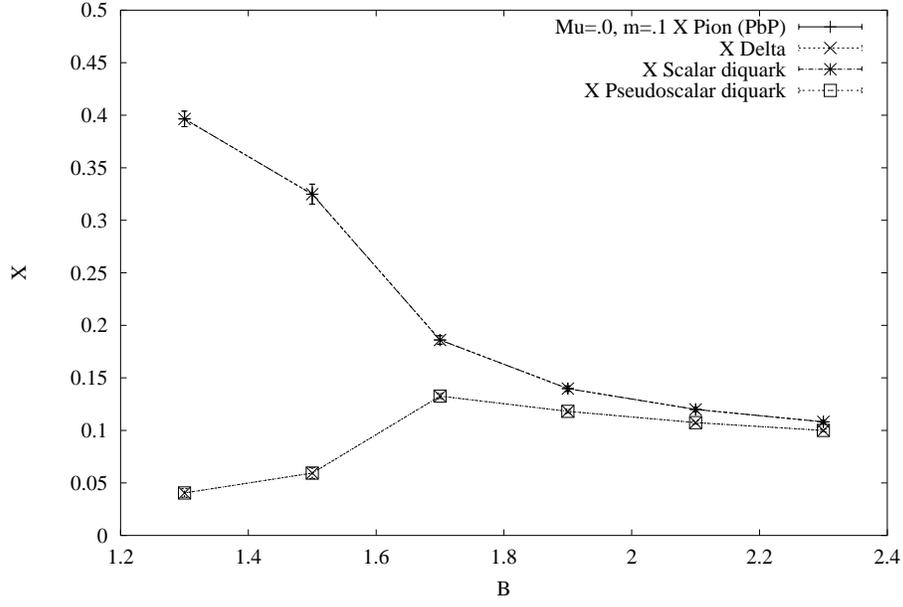, width = 12 truecm}}
    \caption[xxx]{Scalar and pseudoscalar susceptibilieties as
a function of $\beta$ demonstrating chiral symmetry restoration
at high temperature}
    \label{fig:b_I_mass_mu0}
\end{figure}

By increasing the temperature the ``Pauli--G\"ursey'' 
lattice symmetry is preserved, so the above degeneracies
remain true. Chiral condensate is reduced
without changing direction, so the six susceptibilities will
progressively become degenerate, and there will be no residual Goldstone
particle. This behaviour is demonstrated in Fig. \ref{fig:b_I_mass_mu0}
where we plot the susceptibilities as a function of  $\beta$.


The behaviour of the susceptibilities as a function of the
chemical potential has been discussed in HKLM. This is suggestive
of a rotation in chiral space of the condensate, from the
direction ``parallel'' to $<\bar \psi \psi>$ to that ``parallel''
to $<\psi \psi>$.

In HKLM we have noticed that ``conventional'' observables used to
monitor chiral symmetry (pion, $\delta$,
chiral condensate) display a behaviour similar to that observed
at finite temperature. While we confirm this finding at qualitative level,
we also note a quantitative difference : pion and $\delta$ become
exactly degenerate at high density, while at high temperature a
splitting remains -- not surprisingly, of course, since the
bare mass is not zero. It is the apparent {\em  exact} degeneracy
observed for bare masses of comparable value at high density, for both
$\beta$ values, which is sort of puzzling.

\begin{figure}
\begin{center}
{\epsfig{file=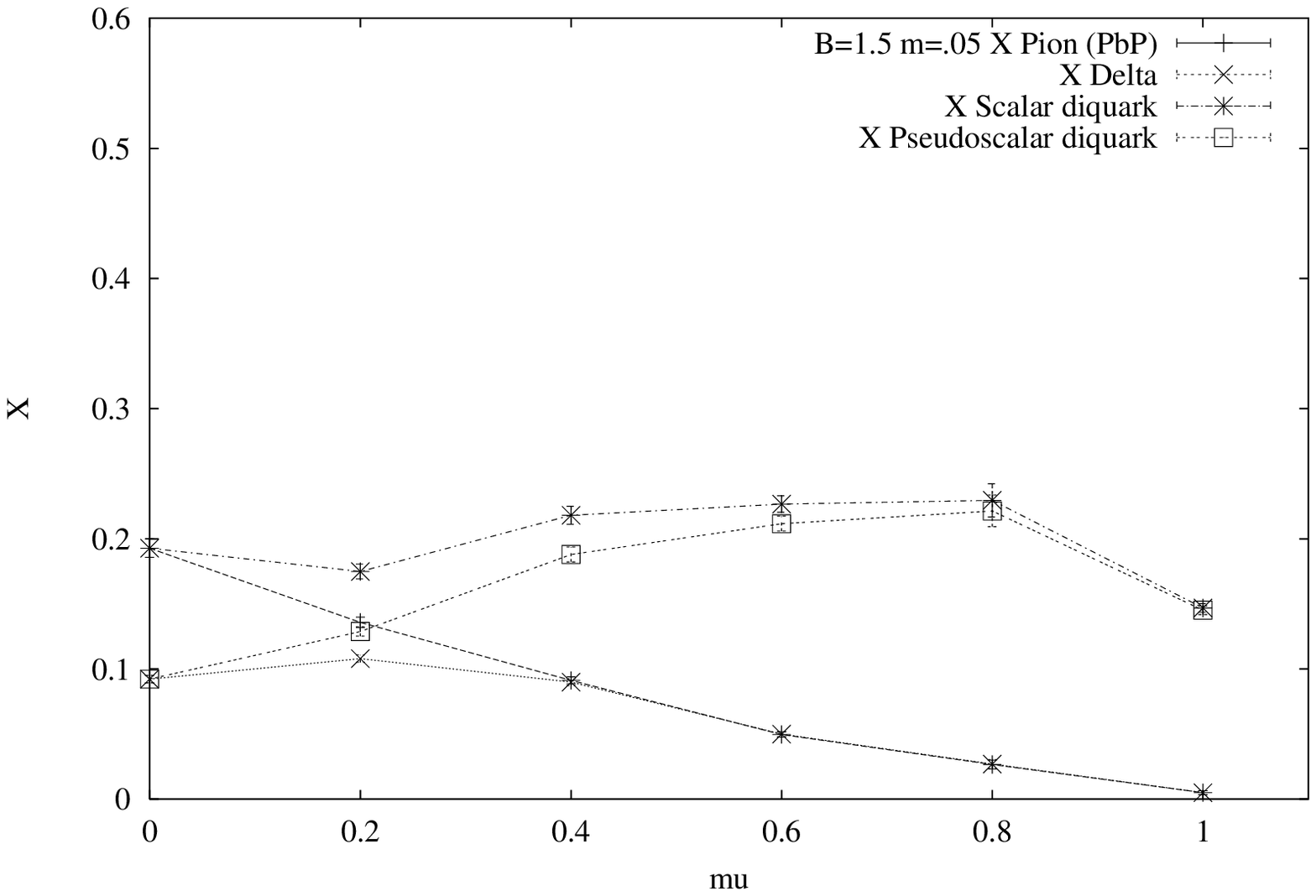, width = 12 truecm}
\epsfig{file=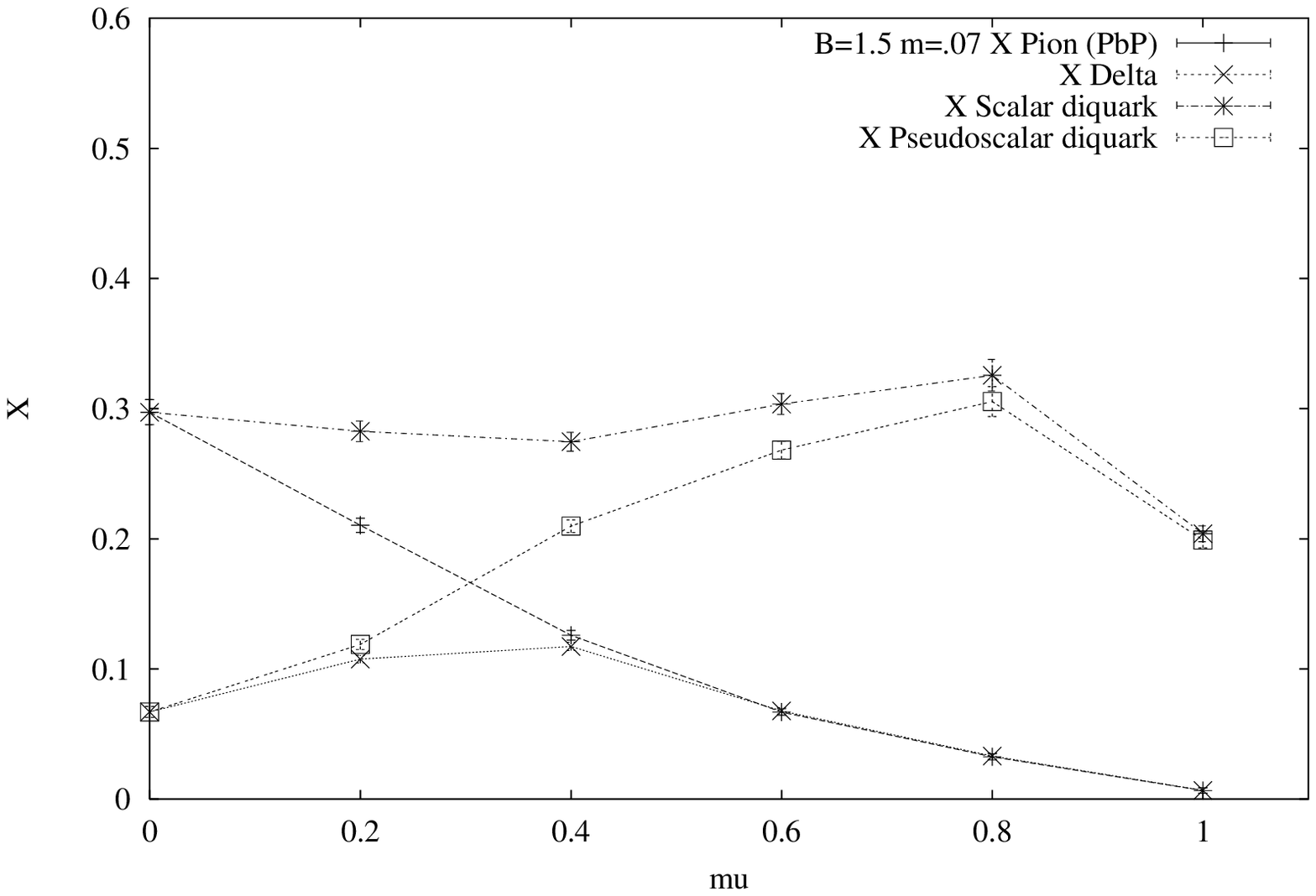 , width = 12 truecm}}
    \caption[xxx]{Scalar and pseudoscalar susceptibilieties 
for mesons and diquarks
a function of $\mu$ at $\beta = 1.5$ and mass = .05
(upper) and mass = .07. See text for discussions}
    \label{fig:mu_I_b1.5}
\end{center}
\end{figure}

\begin{figure}
\begin{center}
{\epsfig{file=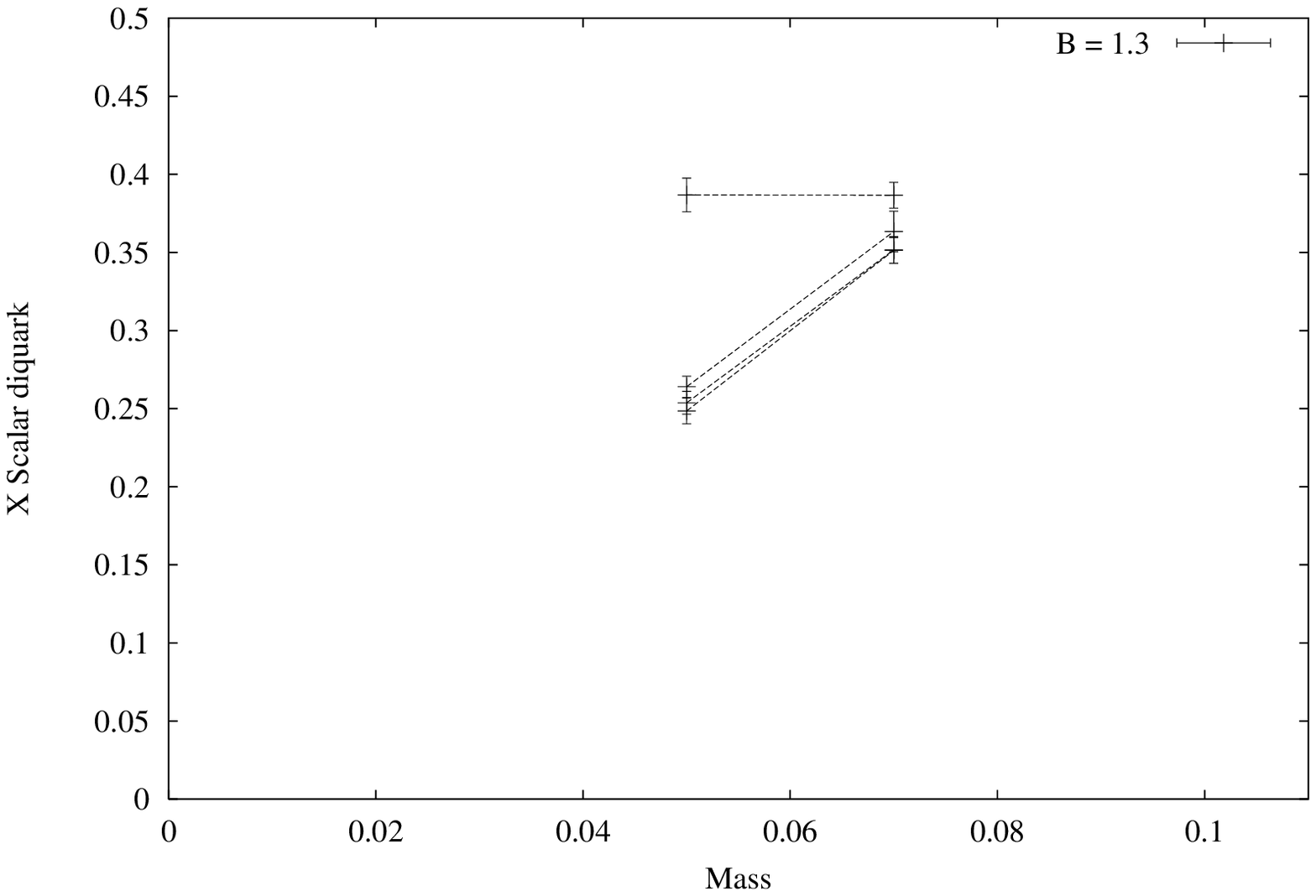, width = 12 truecm}
\epsfig{file=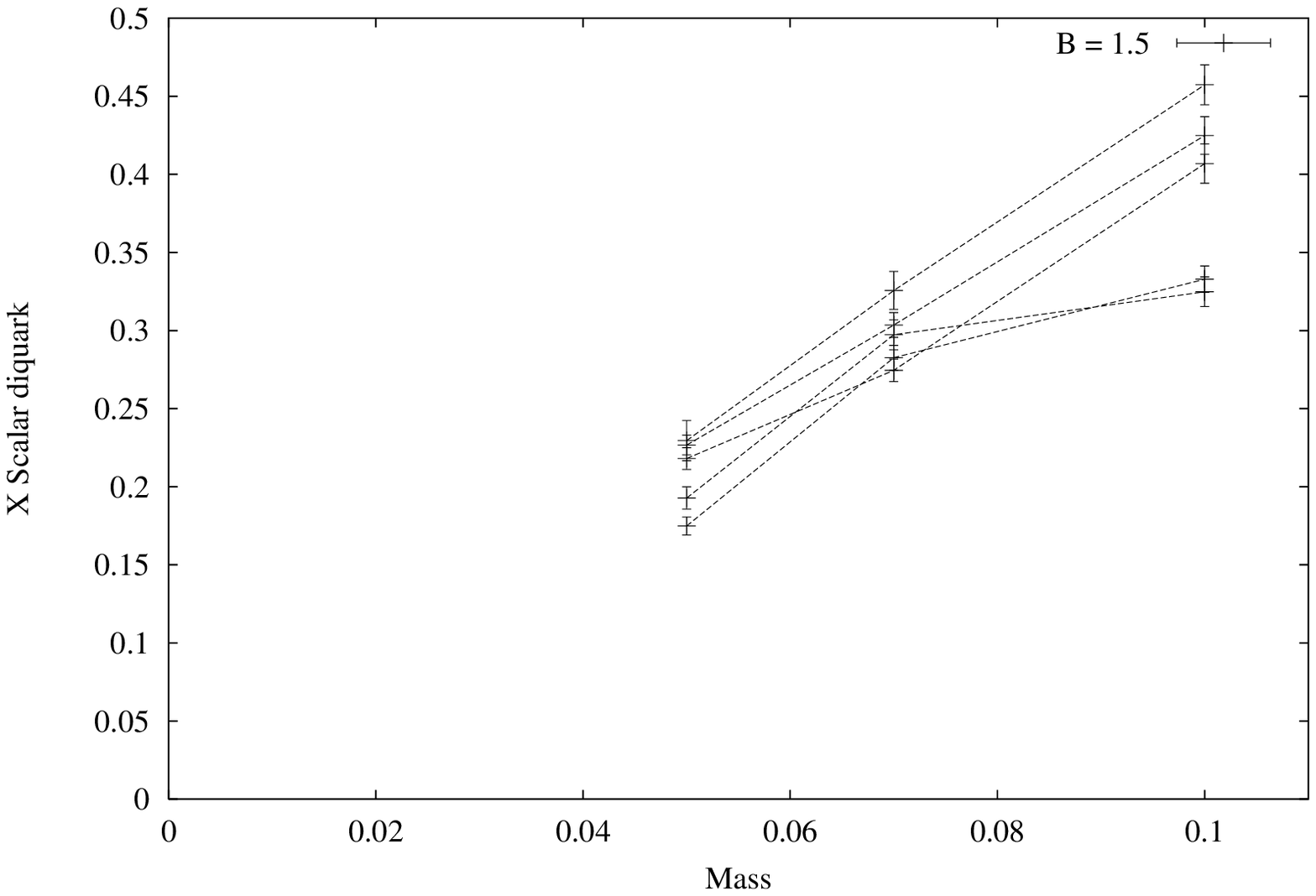, width = 12 truecm}}
\end{center}
    \caption[xxx]{Scalar diquark susceptibilities 
as a function of mass for various $\mu$ at $\beta = 1.3$
and $\beta = 1.5$}
    \label{fig:mu_ibps_mass}
\end{figure}

The qualitative trend observed at $\beta = 1.5$ and $\beta = 1.3$ is
quite similar : see Fig. 
\ref{fig:mu_I_b1.5}, to be contrasted with Fig. 5 of HKLM.
In Figs. \ref {fig:mu_ibps_mass} 
the susceptibility in the scalar diquark channel is plotted as a function
of the bare mass for various 
chemical potentials at $\beta$ = 1.3  and $\beta$ = 1.5.
It looks as though the  signal form the scalar (and pseudoscalar alike)
diquark susceptibilities  in the chiral limit
would be, in any case, much smaller than $<\bar \psi \psi>$ at zero
chemical potential. It has to be stressed that the meaning of this 
exercise is simply
to check how strong the indication  of ``rotation'' detected with this 
procedure would be in  the chiral limit. In addition to that, there
the usual caveats associated with the small lattice size,
and poor mass sample apply. These warnings issued, there might well be
two effects on the chiral condensate while 
increasing $\mu$ : one is indeed a rotation in chiral space, one is a 
reduction of its magnitude similar to that observed at high temperature.

\section{The dynamical effects of the chemical potential : two vs three
colors - speculative comments}
\begin{figure}
\begin{center}
{\epsfig{file=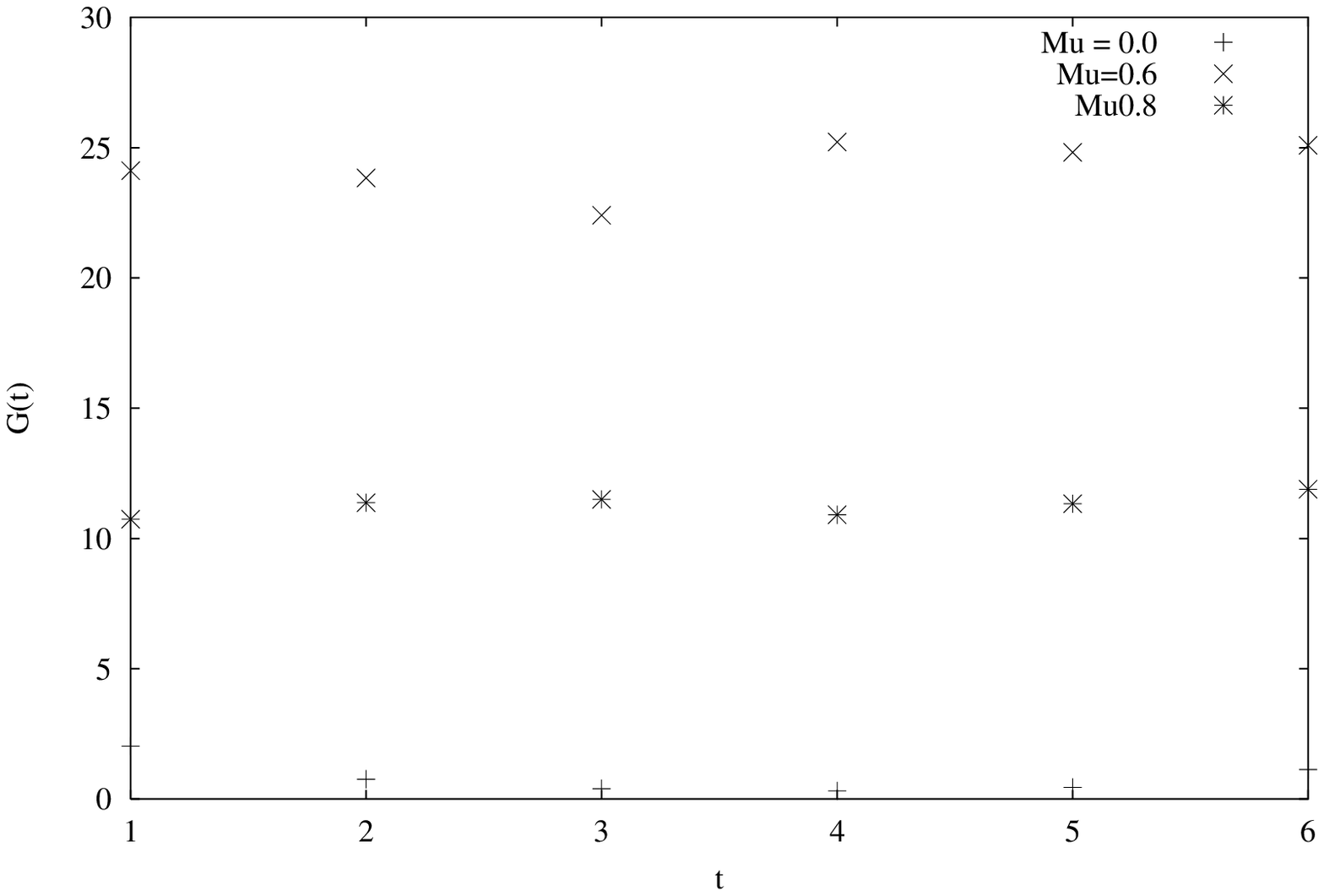, width=12 truecm}
\epsfig{file=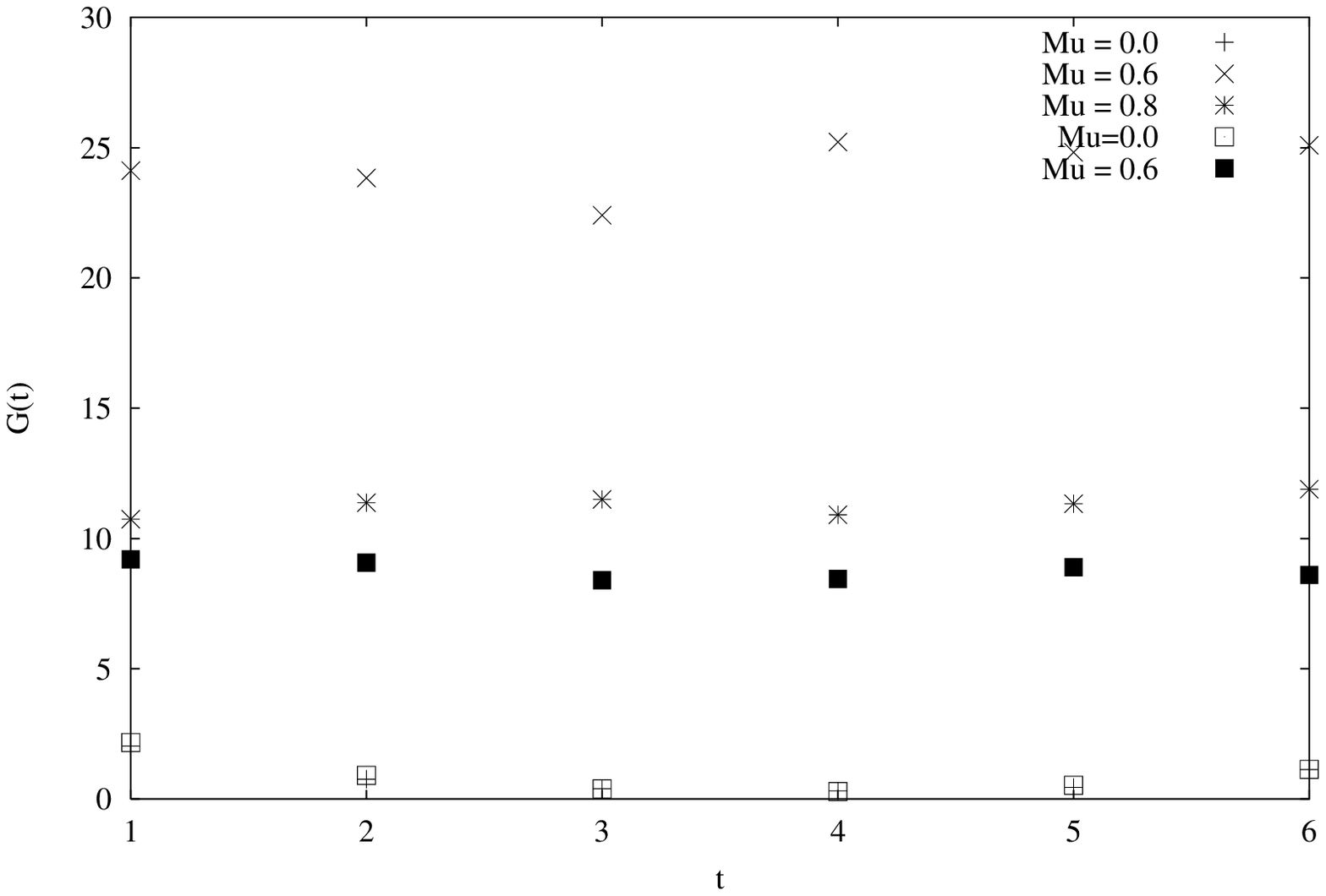, width = 12 truecm}}
\end{center}
    \caption[xxx]{Strong  diquark condensation in the Toy model
on two different configurations.
These two configurations have been generated at $\mu= 0$. The Dirac
propagators have been calculated with 
the chemical potential indicated in the diagrams.}
    \label{fig:Bps_toy_1}
\end{figure}

To gain further insight into the ``dynamical''role of the chemical
potential, and its effect on the gauge fields, we can take a look at a 
Toy \cite {BHKLM} version of the model, obtained by
inverting the Dirac operator at nonzero chemical potential in a
background of gauge fields generated at zero chemical potential.
For these configurations the potential is obviously always ``hard'',
identical to that at $\mu = 0.0$. 
In Fig. \ref{fig:Bps_toy_1} \ we show the scalar diquark
propagator for $\mu = (0, .6 , .8)$ evaluated on two different
configurations. 
Interestingly, we have found that in this case
the behaviour of the diquarks propagators 
resembles that of the infamous quenched SU(3)
``baryonic pions''  measured in \cite{KLS}. 

This suggests that the nature of the interquark forces and
the deconfinement transition might well play
a major role  in the condensation phenomena, and
thus in the critical behaviour, for SU(2) and SU(3) alike.
It is essential to
have the correct quark--quark
and quark--antiquark forces, 
since they  control and soften diquark condensations, including the
pathological ones (which of course should disappear in a correct
SU(3) calculation). In  a correct QCD algorithm
the ``baryonic pion''  condensation would not take place
as the potential become weaker, and the pathological onset
at $m_\pi/2$ would disappear.

\section{Summary and outlook}

Our work in this paper stressed those aspects  which
are instrumental to study the influence of the gauge
field dynamics on the phenomena expected at finite
baryon density.
Temperature dependence is an important tool in this context.
A ``Toy'' model, where the 
gauge field dynamics is not affected by the chemical potential,
proven informative as well.

Although our calculations have not been intended as anything more
than exploratory (remember everything was carried out on a $6^4$
lattice) we feel we have nonetheless managed to learn something
from them, which we summarise here.

We have investigated the interquark potential and found evidence
of enhanced screening when a nonzero  density
is induced in the system, as seen in a larger plateau
at large spatial separation. We underscore that the plateau
builds in is at the onset of the thermodynamics, i.e. when the number density 
starts deviating appreciably from zero. This supports the results
of  heavy quarks SU(2) studies that a small density of quarks induces
``deconfinement'' \cite{heavy2} -- these observations are clearly
semiqualitative and in particular the nature and the very exhistence 
of a nuclear matter state in SU(2) should be subjected
to a more careful scrutiny, as discussed in the body of the
paper.  Temporal and spatial Polyakov loops 
behave in different ways, the latter being nearly insensitive to the 
chemical potential, suggesting the expect different spatial and temporal 
ordering predicted by instanton models. These observations might offer
some further insight into Polyakov Loop 
models \cite{heavy3}.

Simple inspection of the heavy quark potential symmetries 
supports the numerical findings of HKLM, that for any non--zero chemical
potential the (rather heavy) particle degeneracies noted at $\mu=0$ 
remain true, once the Fermi
level shift is taken into account. This would lead to the four degenerate
particles at large chemical potential, when pion and $\delta$, and
scalar and pseudoscalar diquark, are degenerate. Interestingly, this
pattern emerges from a recent model study \cite{modsu2}.

The behaviour of the number density changes qualitatively with temperature.
Despite the many uncertainties described in the text, it seems anyway clear 
that  screening and deconfinement compete against
condensation, and this is better seen on  on a ``warmer'' lattice, close,
yet below $T_c$:  there
$n \propto \mu^3$, consistent with a free, massless quark gas,
suggesting the existence of a critical temperature for diquark 
condensation (i.e. a temperature beyond which diquarks will not condense
at any value of the chemical potential) smaller than $T_c$ itself.

The behaviour of the susceptibilities 
suggests that there are two effects
on the chiral condensate while increasing $\mu$ : one is a rotation
in chiral space, one is a reduction of its magnitude (occuring in the
chiral limit) similar to that we have observed at high temperature. 
The caveats have been spelled out above, and probably only a direct measure of
the diquark condensate \cite {sisu} can clarify this point. 
Assuming it is confirmed, 
there are two possible consequences stemming
from this observation. In the first place, again, this suggests that there
is a critical temperature for diquark condensation: when  temperature
is high enough the magnitude of the condensate will ``shrink'' to zero
at rather small $\mu$,  leading to a complete
realisation of the chiral symmetry $<\bar q q > = < q q> = 0 $.
Secondly, it could be that
this effects ultimately leads to the same complete
restoration of chiral symmetry at
large density even in the cold phase. 

Concerning the comparison of high temperature and high density behaviour,
we noticed two interesting, and to our mind puzzling features:
while at high density we observe a complete degeneracies among the 
chiral partners, at high temperature we observe the residual 
splitting expected of nonzero bare masses; we do not know
where this difference --which is numerically quite clear-- come from.
At the same time, we have not observed any
significant difference between the Polyakov loop correlations
at large T and large $\mu$. Instanton  models predict different
ordering at large T and $\mu$ -- ``instanton-anti-instanton'' pairs
at high T and ``polymers'' at high density (see second entry of
\cite{mod}, and \cite{new})  -- it would
be nice to detect these phenomena in a lattice simulations. 
This would  require more sensitive observables, among which the topological
ones, besides lattice instanton studies, seem to be particularly promising. 
We are now exploring the behaviour of the topological susceptibility, 
and space -- time correlations  of 
topological charge: this should  hopefully shed light on these and other
relevant aspects \cite{topo}. 

We stress that what has been skecthed here
constitute a possible coherent
scenario {\em not inconsistent} with
the present data, but {\em certainly not} implied by it.
Further studies are necessary to sharpen our understanding
and this will certainly require a larger set 
of chemical potentials and simulations in the scaling region.
It is also appropriate to 
reiterate that according to ref. \cite{AS} the theory
with two colors and eight continuum flavour should be in the conformal
phase,  however we have not detected any indication of unconventional
behaviour. Either we are too far from the continuum limit, so that 
the continuum results do not apply, and/or the nature
of this new phase is not manifest in our calculation
\cite {Mira}. One the dark side this issues yet another warning on these 
results, on the bright side this adds to the interest of the model.

A challenging question remains as to whether 
we can learn anything useful for real
QCD. Of course, as  two color QCD baryons are completely 
degenerate with mesons, 
the system is not a good approximation to real QCD -
symmetries and spectrum of  two and three color QCD are 
certainly dramatically different.
However there might well be
a continuity of the dynamical effects from $N_c = 2$ to 
$N_c = \infty$. A good example comes from the behavious of the diquark coupling
\cite {new}. Another from the results of the last section
which suggested that similar dynamical effects should take
place when going from ``Toy'' to full dynamics in two and three color
QCD. In addition, if we accept that the instanton picture 
for ``real'' QCD is correct, and take into account that instantons
should be described as configurations belonging to the SU(2) subgroup
of  SU(3) \cite{adriano}, we might even  speculate that
the two color results would be closer to the real world than expected.

\section*{Acknowledgments}

This paper extends a study initiated in HKLM,
and uses the same codes: I wish to thank Simon Hands, 
John Kogut and  Susan Morrison. I also thank:
the Zentrum f\"ur Interdisziplin\"are 
Forschung, und Fakult\"at f\"ur Physik der Universit\"at Bielefeld; 
the Dipartimenti di Fisica delle Universit\`a 
degli Studi di Pisa e di Roma I {\em La Sapienza}
for hospitality during various stages of this project.
This work was partly supported by 
the  TMR network {\em Finite Temperature Phase Transitions 
in Particle Physics}, EU contract no. ERBFMRXCT97-0122.


\end{document}